# Ordinal Behavior Classification of Student Online Course Interactions

Thomas Trask
CS6795 Spring 2024
Georgia Institute of Technology
thomas.trask@gatech.edu

*Abstract*—The study in interaction patterns between students in on-campus and MOOC-style online courses has been broadly studied for the last 11 years. Yet there remains a gap in the literature comparing the habits of students completing the same course offered in both on-campus and MOOC-style online formats. This study will look at browser-based usage patterns for students in the Georgia Tech CS1301 edx course for both the online course offered to on-campus students and the MOOC-style course offered to anyone to determine what, if any, patterns exist between the two cohorts.

*Keywords—Online Education, Machine Learning, Learning Analytics, EDM, Deep Learning*

## I. Introduction (*Heading 1*)

The advent of Massive open online courses in the last 2 decades has brought about a sea change in how certain students approach their educational opportunities. With free and available access to high-quality content available through MOOCs, the idea was to democratize education for those who most need it. Instead, student access, performance, and retention have been pain points, particularly for those courses in software engineering that scale well via massive access. MOOCs, with their focus on affordability and access, necessitate the use of automated tools that limit opportunities for student-instructor interaction and interplay, an invaluable tool for student success. These and other factors lead to MOOC-style courses often offering canned content and multiple-choice assessments, leading to a perceived/actual degraded student experience for students and disengagement. This, in turn, has led to a lowering of perceived course quality by instructors. Seeking to change that perception Georgia Tech's focuses on offering online parity with some of their on-campus courses. Namely, the CS1301 "Introduction to Computers" class is offered for both on-campus and online students. The qualitative differences between the two offerings are:

- For the on-campus program, students must be enrolled at Georgia Tech. The course is offered on a semester basis and is delivered on a standard 15-week course cadence and is the cost of a standard undergraduate 3-credit course.

- For the MOOC-style program, anyone in good standing with the edX platform is allowed to take the course. The course can be completed in any cadence the student wishes but must be completed by the end of the year the student enrolled in the class in.

Student usage of MOOCs and on-campus courses are well-explored within the literature. However, there is a gap in existing literature when analyzing on-campus and online offerings of the same course. offers a novel opportunity to study usage patterns between the different types of students who may approach each course. This research seeks to explore the following questions:

- What similarities exist in usage patterns between how students approach the different course cadences?
- Can the different cadences be assigned to different usage paradigms?
- Broadly, do pre-existing MOOC usage paradigms persist in on-campus versions of the course?

If there are overlaps between interaction patterns in the disparate courses, interventions developed for one student paradigms may be effective for both cohorts. Additionally, differences in interaction patterns can also allow instructors useful feedback in improving course experiences for each cohort [1,2].

## II. Related Work

### A. Online Education and the importance of effective intervention

Online education has existed since before the world wide web but did not see mass adoption until the advent of the Massive Open Online Course (MOOC) paradigm in 2012 [3,4]. During its development, the creators of the MOOC paradigm had the key goals of "democratizing education" by making access to high-quality education available to anyone with an internet connection, free of charge, and free of barriers [4]. The creators felt that this "field of dreams" approach to education would allow the poorest students to succeed where they otherwise would have failed. In the intervening decade, research has shown a dream deferred [5,6]. Instead of lifting up those students who needed education the most, the technological, social, and academic hurdles needed to overcome to succeed in the type of independent learning MOOCs necessitated, success remained mostly available to those who were already successful: tech-savvy, white-collar men [5,6,7]. Researchers have developed novel tools to help students develop skills in time management, motivation, and continued engagement to help with online school, identifying and providing in-situ interventions for at-risk students, increasing performance, and reducing drop rates [7, 8, 9]. Unfortunately, many novel interventions rely on the built-in temporal and social properties of in-person programs [14, 15, 16].

### B. On-campus and MOOC Interaction Analysis

Because of MOOCs' negligible barriers to entry, they attract a multitude of different students with different goals and expectations. Only 6% of students who sign up for a MOOC course complete the entire breadth of content [5]. For various reasons, roughly 84% of users who sign up for online courses never get past the introductory lesson (70% never open the course page after signing up) [5, 6, 13]. Students that engage with a MOOC's content are most likely finish the course within 2 weeks [10, 11]. Another subset of students engage with a MOOC for a specific subset of the content and report similar satisfaction with the course as those who completed everything.

Because of these different approaches to content consumption, many of the in-situ interventions that have been developed to identify at-risk students in traditional classes ( those operating on a semester-long cadence with temporally-set goals) have shown to be ineffective in predicting MOOC student performance [11]. Recent tools that rely on aggregate content consumption have been shown to be effective in predicting student performance in both MOOCs and on-campus courses [11, 12, 13, CN]. There has been preliminary work done comparing on-campus and on-line versions of the same course using traditional machine-learning methods (SVM, Regression, MLP) analyzing aggregate interaction patterns and achieved positive results [13]. Additional work needs to be done to determine if modern ML algorithms leveraging neural networks can be used.

## III. METHODOLOGY

### A. Course Structure

As mentioned earlier, students take CS1301 through the edX online education platform. The course consists of a series of video lectures, problem sets, and coding problems. The course is split into 4 course sub-modules with the following content breakdown:

TABLE I.  GTX1301 COURSE CONTENT

| Course Sub-module | Videos | Ungraded Exercises | Coding Exercises | Graded Problems |
|---|---|---|---|---|
| **Fundamentals** | 160 | 56 | 54 | 67 |
| **Control Structure** | 122 | 85 | 77 | 85 |
| **Data Structures** | 117 | 58 | 44 | 111 |
| **Objects and Algorithms** | 43 | 17 | 60 | 32 |
| **Totals** | **442** | **218** | **235** | **296** |

Each Sub-course is split into a series of Chapters and chapter sections, grouping similar concepts onto individual pages which the student can consume in any order. There is no requirement that a student finishes full set of content for one section before moving on.

### B. Log Data

Student interaction with the content is logged to a database internally available to Georgia Tech in the form of unstructured log records. Each log record consists of several fields including:
- Who/when/what the interaction was.
- How the user interacted with the content (video interaction duration, problem check score, etc). These will be referred to as 'event types'

This 80M+ record data set is available for 7 instances of the course, 5 on-campus and 2 MOOC instances, organized by year. Of the 39 event types logged, this project will look at those initiated by the user and pertaining to video and problem-check events. Unfortunately, there are no useful log records relating to the ungraded exercises and coding exercises mentioned in table 1. A typical log record has the below structure, with emphasis made to useful fields:

*Figure 1 Example edx Log RECORD*

Log data was filtered to the following user-generated event types:

TABLE 2: EDX LOG EVENT TYPES USED IN ANALYSIS

| Video Event Types | | Problem Check Event Types | |
|---|---|---|---|
| load_video play_video, seek_video, stop_video, | pause_video complete_video hide_transcript speed_change | problem_show problem_graded save_problem_fail problem_check_fail | save_problem_success showanswer problem_check |

## IV. DATA PROCESSING

The course log records were processed in the following main steps to generate an analyzable dataset:

### A. Determine Course Structure

First, the course's structure and content needed to be determined. While the log records contain interaction data, they don't directly contain the course structure and are missing key fields such as content flow and, importantly, titles for any of the videos/static content. To combat this issue, course content was downloaded using the edx-helper data miner and a supplemental data-miner was developed to determine course content [17].

### B. Find Student Interactions

Once the course content path was created, each user's interaction with the above content needed to be determined, i.e. which/how much of each video did the user watch, how well did the users do on the problem sets.

### C. Classify Students based on aggregate interaction patterns

Using the above aggregate interaction patterns, determine the student's interaction ordinal classification. Table 3 details the set of classifications used for determination.

TABLE 3: STUDENT ORDINAL CLASSIFICATION

| Classification | Description | Calculation |
|---|---|---|
| **no-show** | User consumes (close to) no course content | (#_Videos + #_Problems) < 10 |

| Classification | Description | Calculation |
|---|---|---|
| box-checker | User (mostly) only does problem checks | $\#\_Videos/\#\_problems < 10\%$ && $\overline{problem\_attmepts} < 2$ |
| Voyeur | User consumes mostly videos | $\#\_problems/\#\_Videos < 10\%$ && $\#\_Videos > 20$ |
| Studier | User consumes most videos before attempting problem checks & succeeds | $Video\_watchtime > 0.8$ && $problem\_scoreR < 2$ |
| High-engagement | Users engaged significantly with course content | $\overline{watchtime} > 0.8$ & $\overline{ScoreR} < 2$ |
| Normal-engagement | User engaged the standard amount with course content | $\overline{watchtime} > 0.6$ & $\overline{ScoreR} < 3$ |
| Potentially at-risk | User engagement was low, compared to was necessary to complete course content | $\overline{watchtime} > 0.4$ & $\overline{ScoreR} < 4$ |
| At-risk | User engagement was too low | default |

Table 3 references the ScoreR calculation in figure 1 [18]. This calculation is used to compensate for the course allowing users to make unlimited attempts at each problem check; a higher ScoreR value indicates that a user had increasing difficulty passing a particular problem check.

```
if (final_score is passing):
    if # attempts < 2, ScoreR =1
    else if # attempts < 3, ScoreR =2
    elseif # attempts < 5, ScoreR =3
else:
    ScoreR = 4
```
*Figure 2: ScoreR Calculation*

### D. Find common interaction patterns

Using the above classification and aggregate data, determine if there are any overlapping content interaction patterns between students in classification. In addition to aggregate interaction patterns, if time allows the PrefixSpan PySpark interaction pattern analysis package will used to determine what, if any, specific interaction patterns exist between users within a classification.

### V. RESULTS

Before reviewing the results of the research, it is important to reiterate the foundational differences between the course modalities. The on-campus course instances require a user be an active, enrolled student at Georgia Tech. Becoming a GT student is its own feat and generally indicates a base set of technical and academic skills [21]. Conversely, anyone can sign up for the online instances of the course, regardless of skill level, motivation, or ability. While it is not explicitly represented in the log records, it's also important to note that the online course instance can be used as a pre-requisite qualifier for the GT OMSCS program. CS1301 is targeted at users who have little to no experience in writing software.

### A. Enrollment & Engagement

Table 4 shows the enrollment numbers for each course instance, as gathered from the edx logs. As can be seen, the online course instance represents an order of magnitude higher enrollment and engagement compared to the on-campus instances. Additionally, the set of user events for Spring and Summer 2021 are notably smaller, compared to the other courses. This will be discussed in the categorical breakdown.

*TABLE 4 STUDENT ENROLLMENT/ENGAGEMENT*

| Modality | Term | Users | User Events | Sessions |
|---|---|---|---|---|
| On-Campus | Spring 2021 | 341 | 113,016 | 2,120 |
| | Summer 2021 | 94 | 27,849 | 276 |
| | Fall 2021 | 513 | 160,7457 | 4,011 |
| | Spring 2022 | 371 | 1,149,669 | 1,386 |
| | Summer 2022 | 288 | 849,683 | 1,942 |
| | Total | 1607 | 377,675 | 9,735 |
| Online | 2021 | 25670 | 11,833,984 | 79,664 |
| | 2022 | 21346 | 3,916,560 | 24,386 |
| | Total | 47016 | 15,750,544 | 104,050 |

### B. Categorical Breakdown

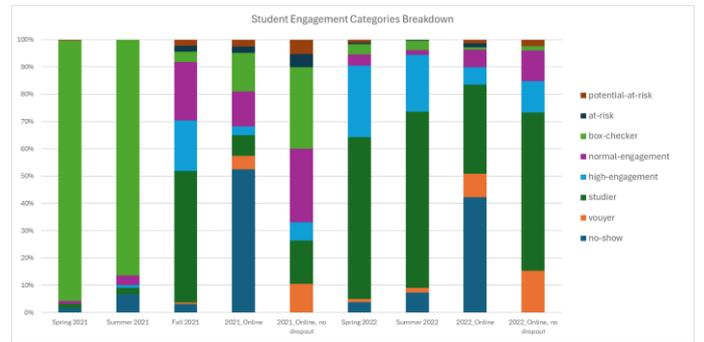

*Figure 3 Student Engagement Categorical Breakdown*

Fig. 3 shows a categorical breakdown for each course instance. There are several things to note about the breakdown.

Spring/Summer 2021 have very few log records indicating that students watched course videos. As such, most of the students were categorized as 'box-checkers'. This may be a deficiency within the log data; a brief temporal analysis of the data indicates that there are sufficient gaps between the different problem check with the course chapters to accommodate for students having watched course videos.

As noted previously, MOOCs often have high dropout rates, which is seen in the chart as both online course instances show a no-show/dropout rate of 40-50%. Charts for each online instance that excluded dropouts were included for easier comparison to on-campus course instances.

In 2022, the vast majority of engaged students in both the on-campus and online course instances qualify as either 'studiers' or 'highly engaged'. While the 2022 on-campus and online cohorts share relatively similar amount of 'studious' users, more online users used the course as a box-check, relative to on-campus students. As noted previously, this may be because of the online course's use as a prerequisite for the OMSCS program.

## C. First and Final scores

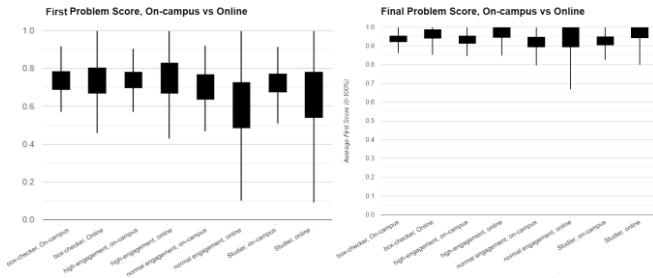

*Figure 4 Problem Check First/final Submission Score, On-campus vs Online.*

Figure 4 shows users average first and final submission score for attempted problem checks. While the first problem's chart shows a significant performance gap between on-campus and online students, the final score chart indicates that that gap closes significantly. On average, both on-campus and online students who engage with the course earn A's. While there is still a gap in performance in the final score's chart, it may be explained by the broader cohort comprising the online campus.

## D. ScoreR

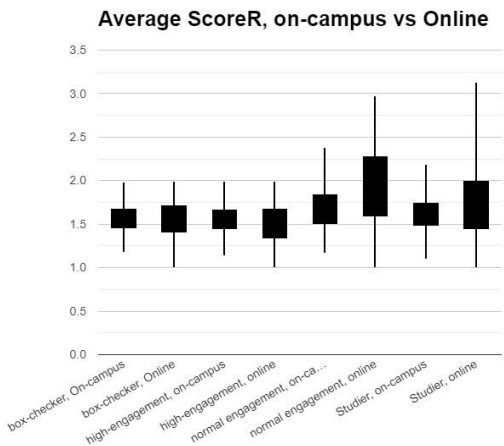

*Figure 5 Average ScoreR, On-campus vs Online*

As discussed earlier, ScoreR is a useful tool in scaling a student's effort relative to their performance on problem checks [18]. Fig. 3 shows the variance in ScoreR for each major student category for both on-campus and online students. Of note is the large difference in variance between on-campus and online instances of the same category. This may be an indicator of the base set of study/technical skills that GT students develop, as previously mentioned. There are other explanations for this result, such as variances in language, general unfamiliarity with the course content, or a wider demographic representation than that found at GT.

## E. Temporal Engagement

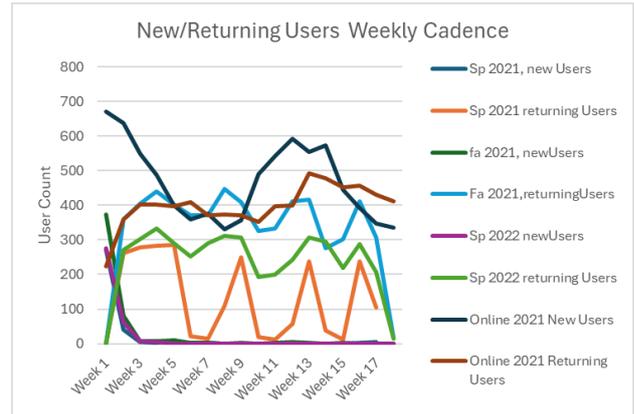

*Figure 6 Per-Week New/Returning counts*

Figure 6 shows On-campus and MOOC students approaching course content at different cadences. While on-campus students split their time evenly week-to-week, engaged online students complete the course within a 2-week period. While the MOOC course is offered year-round, student engagement tends to peak around the beginning of the Fall and Spring semesters (to "skill up" for upcoming courses), and the end of each year due to user's course progression resetting at the end of each calendar year. Figure 6 shows the number of new users for each on-campus course dropping to 0 after the first week of a course but shows a sustained amount of new and returning users for the online course.

## F. Fine-grain Interaction Patterns

As part of this project, a tool leveraging the Pyspark PrefixSpan interaction pattern analysis package was developed. While it did find general trends of interactions (most students who are unsuccessful answering a problem-check will not return to course videos before trying the problem again), it was not useful in determining fine-grain interaction patterns for any users, regardless of category. Additionally, the PySpark library, which runs through the JVM, was grossly unstable and slow during development. More time should be given to develop the tool to determine if any fine-grain interaction patterns exist within user categories.

## VI. CONCLUSION

### A. Categorical Representation

Categorical Representation appears to be successful in determining usage patterns between different student paradigms. Student paradigms that have been identified in MOOCs (high/normal-engagement, vouyer, box-checker) are all present in the on-campus courses to some extent. The relative representational consistency in the 3 2022 course instances and consistent box-plot margins across multiple instances seems to indicate that the calculations for each are at least semi-accurate. Being able to identify a student's usage paradigm based on log data can be a huge win for implementing targeted interventions. To wit: the performance gap in the 'studious' users in both modalities would be a useful target for intervention as they spend the most time in the course and receive the widest grade distribution. While almost all engaged students ended up achieving high class scores, 'studious' students showed the least consistent final score. Struggling students that are putting in work may benefit from supplemental help or help in other non-school related skills such as effective studying habits or time management.

*B. MOOC Motivations*

As previously noted, students approach MOOCs from several motivations. Some use them to dive deep into an unfamiliar subject (studiers). Some use them as a stepping-stone to greater heights (box-checkers) [2, 3]. Some use them for specific content and move on (voyeurs). Significant Skill gaps can be seen between students in the different modalities that may indicate the above varying motivations but may also represent a need for interventions.

## VII. LIMITATIONS AND CONSIDERATIONS

As noted previously, there are key limitations to this work. First, CS1301 students are given infinite opportunities to complete assessments. As a result, students who complete all course content generally also end up with full marks on assessments. This has been addressed somewhat by examining student's first scores and their Score R. It has been noted that many students 'brute-force' their way through the course problem checks, trying a problem until they achieve a passing score. There seems to be a gap in the usage data for Spring and Summer 2021 where most users show no record of watching course videos. The data being accessed is via a historical log database and is at least a year old. Edx changes course and log formatting within/between course instances which can make the data difficult to process. The linking used to determine user interaction flow was not directly available via either the log dataset or the GT edx database which may cause inconsistencies within the flow.

Many things were not available through the edx log, such as user demographic, survey data, programming assignments/grades, ungraded assessment submission/grades, class forum activity (either through Piazza, EdStem, or Canvas). Course syllabi was also not available. While the categories used were based on existing calculations, a deeper analysis of the student data should be done to ensure that they accurately represent student interactions. The dataset only represents 7 instances of one course and the analysis done here is thus subject to selection bias.

## VIII. FUTURE WORKS

This project stands as only the second foundational work in the GT edx data processing pipeline. The edx log data is one piece of a very large student profile. Additional work may involve linking student data to class forum content and other courses to track a student's long-term progress. Courses from a wide breadth of subjects need to be analyzed to determine how well the student categories translate across courses. Forum posts can be analyzed via LLM to determine a student's preferred language and make content translations available. Fine-grain video access analysis can be done to determine what parts of a video students struggle with, which can allow for content improvement. Once a student's ordinal learning style is determined, edx Learning Tool Integrations (LTIs) can be developed to allow bespoke in-situ interventions. As previously mentioned, the PrefixSpan tool should be further developed to allow for detection of fine-grain interaction patterns within user categories.

## IX. REFERENCES

[1] Tawfik, A. A., Shepherd, C. E., Gatewood, J., & Gish-Lieberman, J. J. (2021). First and second order barriers to teaching in K-12 online learning. TechTrends, 65(6), 925-938. https://doi.org/10.1007/s11528-021-00648-y

[2] Wise, A. F. (2014, March). Designing pedagogical interventions to support student use of learning analytics. In Proceedings of the fourth international conference on learning analytics and knowledge (pp. 203-211).

[3] Kentnor, H. E. (2015). Distance education and the evolution of online learning in the United States. Curriculum and teaching dialogue, 17(1), 21-34.

[4] Pappano, L. (2012). The Year of the MOOC. The New York Times, 2(12), 2012.

[5] Yang, Diyi; Sinha, Tanmay; Adamson, David; Penstein Rose, Carolyn (2013) Turn on, tune in, drop out: Anticipating student dropouts in massive open online courses in Neural Information Processing Systems: Workshop on Data-Driven Education (NIPS 2013), Lake Tahoe, NV, USA, December 9-10, 2013

[6] Xiong, Y., Li, H., Kornhaber, M.L., Suen, H.K., Pursel, B. & Goins, D.D. (2015). Examining the relations among student motivation, engagement, and retention in a MOOC: A structural equation modeling approach. Global Education Review, 2 (3). 23-33 https://files.eric.ed.gov/fulltext/EJ1074099.pdf

[7] Kizilcec, René F.; Halawa, Sherif A (2015) Attrition and Achievement Gaps in Online Learning in Proceedings of the Second (2015) ACM Conference on Learning @ Scale, March 2015 https://rene.kizilcec.com/wpcontent/uploads/2013/02/kizilcec2015attrition.pdf

[8] Zhang, Q., Peck, K.L., Hristova, A. et al. Exploring the communication preferences of MOOC learners and the value of preference-based groups: Is grouping enough?. Education Tech Research Dev 64, 809–837 (2016).

[9] Gardner, J., Brooks, C. Student success prediction in MOOCs. User Model UserAdap Inter 28, 127–203 (2018). https://arxiv.org/pdf/1711.06349.pdf

[10] Borrella, I., Caballero-Caballero, S., & Ponce-Cueto, E. (2022). Taking action to reduce dropout in MOOCs: Tested interventions. *Computers & Education*, 179, 104412.

[11] Bote-Lorenzo, Miguel L.; Gómez-Sánchez, Eduardo (2017) Predicting the decrease of engagement indicators in a MOOC in *Proceedings of the Seventh International Learning Analytics & Knowledge Conference* March 2017 Pages 143–147

[12] Chiu, Y. C., Hsu, H.-J., Wu, J., & Yang, D.-L. (2018). Predicting student performance in MOOCs using learning activity data. Journal of Information Science and Engineering, 34, 1223-1235. https://doi.org/10.6688/JISE.201809_34(5).0007

[13] Vostatek, V. C. (2020). Predicting factors that affect student performance in MOOC and on-campus computer science education (Doctoral dissertation, Massachusetts Institute of Technology).

[14] Liu, M., McKelroy, E., Corliss, S. B., & Carrigan, J. (2017). Investigating the effect of an adaptive learning intervention on students' learning. Educational technology research and development, 65, 1605-1625.

[15] Fernandez-Rio, J., Sanz, N., Fernandez-Cando, J., & Santos, L. (2017). Impact of a sustained Cooperative Learning intervention on student motivation. Physical Education and Sport Pedagogy, 22(1), 89-105.

[16] Outhwaite, L. A., Gulliford, A., & Pitchford, N. J. (2020). A new methodological approach for evaluating the impact of educational intervention implementation on learning outcomes. International Journal of Research & Method in Education, 43(3), 225-242.

[17] Zheng, Ye, (2023) E*dx-helper*, Github repository, https://github.com/csyezheng/edx-helper

[18] Mubarak, A. A., Cao, H., Hezam, I. M., & Hao, F. (2022). Modeling students' performance using graph convolutional networks. Complex & Intelligent Systems, 8(3), 2183-2201.

[19] Hu Q, Rangwala H (2019) Academic performance estimation with attention-based graph convolutional networks. In: EDM 2019—Proceedings of the 12th international conference on educational data mining. international educational data mining society, pp 69–78

[20] Crick, F., & Koch, C. (2007). A neurobiological framework for consciousness. The Blackwell companion to consciousness, 567-579.

[21] PrepScholar, "Georgia Tech SAT & GPA Scores", Reviewed on April 2, 2024, https://www.prepscholar.com/sat/s/colleges/Georgia-Tech-sat-scores-GPA